\documentclass{PoS}

\title{Photon-ALP conversions inside AGN}

\ShortTitle{Photon-ALP conversions inside AGN}

\author{\speaker{Fabrizio Tavecchio}\\
       INAF-Osservatorio Astronomoco di Brera in Merate, Merate, Italy\\
       E-mail: \email{fabrizio.tavecchio@brera.inaf.it}}

\author{Giorgio Galanti\\
        Dipartimento di Fisica, Universit\`a dell'Insubria, Via Valleggio 11, I -- 22100 Como, Italy\\
        E-mail: \email{gam.galanti@gmail.com}}

\author{Marco Roncadelli\\
        INFN, Sezione di Pavia, Via A. Bassi 6, I -- 27100 Pavia, Italy\\
        E-mail: \email{marco.roncadelli@pv.infn.it}}

\abstract{An intriguing possibility to partially circumvent extragalactic background light (EBL) absorption in very-high-energy (VHE) observations of blazars is that photons convert into axion-like particles (ALPs) $\gamma \to a$ inside or close to a blazar and reconvert into photons $a \to \gamma$ in the Milky Way magnetic field. This idea has been put forward in 2008 and has attracted a considerable interest. However, while the probability for the back-conversion $a \to \gamma$ has been computed in detail (using the maps of the Galatic magnetic field), regretfully no realistic estimate of the probability for the conversion $\gamma \to a$ inside a  blazar has been performed, in spite of the fact that the present-day knowledge allows this task to be accomplished in a reliable fashion. We present a detailed calculation that fills this gap, considering both types of blazars, namely BL Lac objects (BL Lacs) and flat spectrum radio quasars (FSRQ) with their specific structural and environmental properties. We also include the host elliptical galaxy into account. Our somewhat surprising results show that the conversion probability in BL Lacs is strongly dependent on the source parameters -- like the position of the emission region along the jet and the strength of the magnetic field therein -- making it effectivelly {\it unpredictable}. On the other hand, the lobes at the termination of FSRQ jets lead to an effective ``equipartition" between photons and ALPs due to its chaotic nature, thereby allowing us to make a clear-cut prediction. These results are quite important in view of the planned VHE detectors like the CTA, HAWK and HiSCORE.}

\FullConference{Science with the New Generation of High Energy Gamma-ray experiments, 10th Workshop \\
		 04-06 June 2014\\
		 Lisbon - Portugal}

\begin{document}

\section{Introduction}    

Axion-like particles (ALP) are spin-zero, neutral and very light new particles denoted by $a$ throughout (for a review, see~\cite{alps}),  generically predicted by many extensions of the Standard Model of particle physics and in particular by superstring theories (for a review, see~\cite{ringwald}). Basically they can be regarded as a generalization of their archetype, the axion, which is the pseudo-Goldstone boson arising from the breakdown of the global  Peccei-Quinn symmetry $U(1)_{\rm PQ}$ invoked as a natural solution of the well known strong CP problem. But in order to allow the analysis to be as much model-independent as possible, two assumptions are made: (1) {\it only} the two-photon coupling $a\gamma\gamma$ is considered (discarding any possible coupling to fermions and gluons) and (2) the mass $m$ and the two-photon coupling constant $1/M$ are {\it independent}. Accordingly, the ALP Lagrangian is 
\begin{equation}
\label{t1}
{\cal L}^0_{\rm ALP} = \frac{1}{2} \, \partial^{\mu} a \, \partial_{\mu} a - \, \frac{1}{2} \, m^2 \, a^2 + \frac{1}{M} \, {\bf E} \cdot {\bf B} \, a~,
\end{equation}
where for our purposes ${\bf E}$ denotes the electric field of a propagating photon and ${\bf B}$ is an {\it external} magnetic field. Because of the structure of the interaction term ${\bf E} \cdot {\bf B} \, a$, only the component of $\bf B$ parallel to the photon polarization couples to $a$: due to the fact that the photon polarization is transverse to its momentum, such a $\bf B$ component is called transverse and denoted by ${\bf B}_T$. Having assumed that ${\bf B}$ is an external field, the interaction term $E \, B_T \, a$ gives rise to the $\gamma a$ mixing, which implies that photon-ALP conversions $\gamma \to a$ and $a \to \gamma$ -- as well as photon-ALP oscillations $\gamma \leftrightarrow a$ -- occur~\footnote{Photon-ALP oscillations are quite similar to oscillations involving massive neutrinos of different flavors, apart from the difference that here an external ${\bf B}$ field is necessary in order to compensate for the spin mismatch between photons and ALPs.}. The only robust bound on $M$ is provided by the CAST experiment which gives $M > 1.14 \cdot 10^{10} \, {\rm GeV}$ for $m < 0.02 \, {\rm eV}$~\cite{cast}.

For situations involving relatively large magnetic fields and/or photon energies, the $\gamma a$ mixing effect tends to be offset by the one-loop QED vacuum polarization effect, which has therefore to be taken into account and is described by the Lagrangian~\cite{HEW}
\begin{equation}
\label{t1q}
{\cal L}_{\rm HEW} = \frac{2 \alpha^2}{45 m_e^4} \, \left[ \bigl({\bf E}^2 - {\bf B}^2 \bigr)^2 + 7 \bigl({\bf E} \cdot {\bf B} \bigr)^2 \right]~,
\end{equation}
where $\alpha$ is the fine-structure constant and $m_e$ is the electron mass. So, the full ALP Lagrangian we are concerned with is 
${\cal L}_{\rm ALP} = {\cal L}^0_{\rm ALP} + {\cal L}_{\rm HEW}$. If the conversion occurs in a background plasma, it is necessary to include also plasma effects, which are formally taken into account by a photon mass equal to the plasma frequency 
$\omega_{\rm pl}$~\cite{rs1988}. Before turning to a different issue, it proves convenient in view of our subsequent discussion to define a low-energy threshold $E_L$ and an high-energy one $E_H$ as
\begin{equation}
\label{ETS}
E_L \equiv \frac{M | m^2 - \, \omega^2_{\rm pl} |}{2 \, B_T}~, \ \ \ \ \ \ \ \ \ \ \ E_H \equiv \frac{45 \pi}{3.5 \alpha} \left(\frac{B_{\rm cr}}{B_T} \right)^2 \left(\frac{B_T}{M} \right)~,
\end{equation}
where $B_{\rm cr} = 4.41 \cdot 10^{13} \, {\rm G}$ is the critical magnetic field. Then the condition $E_L < E < E_H$ defines the {\it strong-mixing regime}, in which the photon-ALP conversion probability $P_{\gamma \to a} (E)$ becomes maximal and 
energy-independent\footnote{When ${\bf B}$ is position-dependent, also $E_L$ and $E_H$ become position-dependent and the same is true for the strong-mixing condition.}. We anticipate that in general we will be outside the strong-mixing regime.\\

The importance of photon-ALP conversions and oscillations has been widely discussed in the framework of high ($E > 100 \, {\rm MeV}$) and very-high ($E > 100 \, {\rm GeV}$) extra-galactic astrophysics, where the great majority of sources are {\it blazars}, namely active galactic nuclei (AGN) characterized by the presence of a relativistic jet of plasma closely aligned with the observer's line of sight~\cite{urry}. Specifically, most of the work has addressed ALP effects on the observed $\gamma$-ray spectrum in a few situations where a magnetic field is always present: $\gamma \leftrightarrow a$ oscillations in extragalactic space~\cite{darma,dgr2011}, $\gamma \to a$ conversions in the source and $a \to \gamma$ reconversions in the Milky Way~\cite{hooper2008,wb,harris}, $\gamma \to a$ conversions in a galaxy cluster when the source is embedded in it and $a \to \gamma$ reconversions in the Milky Way~\cite{cluster} and finally a combination of the first two scenarios~\cite{prada}. 

So far, nobody has correctly estimated the $\gamma \to a$ conversion probability $P_{\gamma \to a} (E)$ in blazars. This fact motivates our work presented below, indeed aimed at evaluating $P_{\gamma \to a} (E)$ in blazars using the most updated physical information concerning their magnetized regions.

After a discussion of our knowledge of the jet physical properties derived by current observations and modeling (\S 2) we derive $P_{\gamma \to a} (E)$ for both BL Lacs and FSRQ (\S 3), and finally we discuss the results (\S 4). A preliminary account of the matter presented here has been reported in~\cite{letter}, whereas a much more detailed analysis will appear in two forthcoming papers. 

\section{Setting the stage}

Rather schematically, the current view of blazars can be sketched as follows~\cite{dermer14}. Their central engine is a rotating (Kerr) supermassive black hole (SMBH), with mass often exceeding $10^8$ solar masses, lying at the centre of an elliptical galaxy and accreting matter from the surrounding. The infalling material heats up before disappearing into the SMBH, thereby giving rise to the emission of an enormous amount of {\it thermal} radiation concentrated in the optical-ultraviolet band. Not well understood processes  (driven by magnetohydrodynamic effects mediated by the magnetic field supported by the accretion flow) give rise to two oppositely oriented jet configurations originating from the central SMBH~\cite{simul}. What happens is that while the first part of the jet, still subject to an outward acceleration, is basically dissipationless, when it reaches its asymptotical speed -- at about $10^{16} - 10^{17} \, {\rm cm}$ from the SMBH -- internal shocks and/or magnetic reconnection can occur, leading to the acceleration of electrons. As  consequence, they emit beamed synchrotron and inverse Compton radiation, producing the {\it non-thermal} radiation that we observe. This radiation is strongly boosted in the direction of the plasma bulk speed by relativistic aberration: typical bulk Lorentz factors of the jets are around $\Gamma=10-20$, hence implying that the radiation is strongly beamed within a cone with semi-aperture $\theta \simeq 1/\Gamma \simeq 2-5$ degrees. As a matter of fact, a double-humped SED is produced, with the first peak lying somewhere between the infrared and the X-ray band which is due to the synchrotron emission of relativistic electrons in the jet, while the second peak lies in the $\gamma$-ray band. However, the origin of the latter peak is debated. Two mechanisms have been proposed for its origin: one leptonic and the other hadronic. In the leptonic case~\cite{lept} the peak is due to the inverse Compton (IC) scattering off the same electrons responsible for the synchrotron peak (with a possible contribution from external photons), while in the hadronic mechanism~\cite{hadr} the considered peak is due to reactions involving relativistic hadrons with neutral and charged pions decaying into $\gamma$-rays and neutrinos, respectively. Assuming a standard one-zone emission model~\cite{gg10}, robust values for the basic parameters like the magnetic field ${\bf B}$, the electron number density $n_e$ and the size of the emission region ${\cal R}_{{\rm ER}}$ are derived. Unfortunately, some degree of uncertainty still remains, in particular concerning the location $d_{{\rm ER}}$ of ${\cal R}_{{\rm ER}}$ along the jet and the strength and geometry of ${\bf B}$. Beyond ${\cal R}_{{\rm ER}}$, the produced $\gamma$-rays travel along the jet. Denoting by $d$ the coordinate along the jet axis as measured from the centre, ${\bf B} (d)$ can be probed through radio polarimetric techniques that allow us to determine its geometry and intensity. These studies are more conclusive for BL Lac jets, for which a consistent view is emerging. For FSRQ jet the situation is more complex, since the jets do not display a common behavior. In the following we shall discuss separately the two cases.

\subsection{BL Lacs}

State-of-the art modeling of BL Lacs emitting at VHE~\cite{tavecchio2010} provides the values of the main physical quantities at 
${\cal R}_{{\rm ER}}$, $B (d_{{\rm ER}}) = 0.1 - 1 \, {\rm G}$ and $n_e (d_{{\rm ER}}) \simeq 5 \cdot 10^{4} \, {\rm cm}^{-3}$. The determination of $d_{{\rm ER}}$ itself is difficult to get directly. Indirect estimates based on the size of ${\cal R}_{{\rm ER}}$ and under the hypothesis of a conical jet geometry yield $d_{{\rm ER}} \simeq 10^{16} -10^{17} \, {\rm cm}$.

Beyond ${\cal R}_{{\rm ER}}$ photons travel outwards unimpeded until they leave the jet and propagate into the host galaxy. Due to the limited instrumental sensitivity, most of the studies of jet structure focus on the brightest knots at parsec and multi-parsec distance from the centre, believed to flag internal and/or external shocks in the flow. The magnetic field ${\bf B} ( d )$ can be decomposed into a {\it toroidal} part transverse to the jet axis $B_T ( d ) \propto d^{- 1}$ and a {\it poloidal part} parallel to the jet axis $B_P ( d ) \propto d^{- 2}$~\cite{bbr1984}. At large distances from the SMBH, the toroidal component evidently dominates.

This view is supported by recent work that has succeeded to observationally characterize the magnetic field structure over distances in the range $0.1 - 100 \, {\rm pc}$ in several jets of BL Lacs  through polarimetric studies (see e.g.~\cite{pudritz2011} and references therein). These works convincingly demonstrate that in BL Lacs the magnetic field is substantially {\it ordered} and predominantly {\it traverse} to the jet axis. This is clearly observed in the brightest knots of the jet~\cite{asada2008}, but there are indications that the same geometry is associated also to the intra-knot regions~\cite{gabuzda1999}, supporting the view that the presence of a well-ordered, transverse field component is a structural characteristic of these jets.  Note that, consistently with this framework, the inferred intensity of ${\bf B}$ is observed to scale as the inverse of the distance along the jet~\cite{osullivan2009}. We also remark that these results rule out any domain-like structure of the magnetic field in the jet (as assumed e.g. in~\cite{harris,prada}).

Hence, supported by these arguments we assume that the magnetic field is ordered and transverse to the jet axis for $d > 
d_{{\rm ER}}$, and that its strength is
\begin{equation}
\label{Bjet}
B_T ( d ) = B_T (d_{{\rm ER}}) \left(\frac{d}{d_{{\rm ER}}} \right)^{- 1}~. 
\end{equation}
Observe that in Eq. (\ref{Bjet}) ${\bf B}_T$ is measured in a co-moving frame with the jet plasma, namely with the Lorentz factor 
$\Gamma$ with respect to a stationary observer. The transformation to the stationary frame is then simply performed by means of the replacement $E \to \Gamma E$ in the final result. Under the usual assumption that the jet has a conical shape we expect that the electron number density is
\begin{equation}
\label{nejet}
n_e ( d ) = n_e (d_{{\rm ER}}) \left( \frac{d}{d_{{\rm ER}}} \right)^{- 2}~,
\end{equation}
which again holds true in the jet co-moving frame.

Of course, in real jets, these smooth profiles, holding when large scales are considered, are likely disturbed by the presence of shocks and/or other flow instabilities. Such variations, occurring on relatively small scales, could lead to modifications of the 
total $P_{\gamma \to a} (E)$ that we are going to compute, possibly introducing distortions in the resulting ALPs (or photon) spectra, but this does not affect our main findings. 

An important parameter is the length of the jet, fixing the characteristic length scale where the $\gamma \to a$ conversions can occur in the jet magnetic field. Observations suggest that the majority of the jets associated with TeV BL Lacs are relatively short, losing their collimation and coherence at distances of the order of $1 \, {\rm  kpc}$~\cite{giroletti2004}. 

\subsection{FSRQs}

FSRQs present some additional complications with respect to the simpler case of BL Lacs. Within a distance $d_{{\rm BLR}} \simeq10^{18} \, {\rm cm}$ from the centre the jet is surrounded by the radiation emitted by the clouds occupying  the so called {\it broad line region} (BLR). At larger distances the importance of the BLR field decreases and the external regions are dominated by the IR field of a  dusty torus reprocessing part of the radiation emitted by the central accretion flow. The photons belonging to these external fields can interact with the gamma rays through the process $\gamma \gamma \to e^+ e^-$, disappearing from the beam. In the absence of any conversion, the decrease of the intensity of the beam is generally given by $I(E)\propto \exp[-\tau_{\gamma\gamma}(E)]$, where $\tau_{\gamma\gamma}(E)$ is the energy-dependent optical depth, function of the target photon density and distance. Detailed calculations \cite{1222} show that the BLR is practically opaque for energies $E > 20 \, {\rm GeV}$, while the dusty torus provide substantial absorption above $\sim 1$ TeV. Since part of the photons convert to ALP -- which are not absorbed because they do not interact with anything~\footnote{Based on the interaction term in ${\cal L}^0_{\rm ALP}$, it is straightforward to get the following order-of-magnitude estimate for the corresponding cross-sections $\sigma (a \, \gamma \to f^+ f^-) \sim \sigma (a \, f^{\pm} \to \gamma \, f^{\pm})  \sim 10^{- 52} \, {\rm cm}^2$, where $f$ denotes any charged fermion.} -- the effective optical depth can be smaller than in the conventional case, possibly explaining the puzzling detection of few FSRQ at VHE~\cite{1222}.

For the VHE $\gamma$-ray emission region  we take $d_{\rm ER}$ larger by a factor of 3 as compared to the BL Lac case, based on the larger variability time scales. The modeling of the SED provides $B_T (d_{{\rm ER}}) = 1 - 5 \, {\rm G}$ and $n_e (d_{{\rm ER}})  \simeq 10^4 \, {\rm cm}^{- 3}$~\cite{gg10}. The geometry and the intensity of ${\bf B}$ in the jet beyond ${\cal R}_{{\rm ER}}$ are far less clear than in the case of BL Lacs. In fact, there are indications that ${\bf B}$ has a globally ordered structure, but its inclination angle $\varphi$ with respect to the jet axis does not have a unique value for all sources, actually covering the whole interval $0 - 90^{\circ}$. For definiteness, we assume the same profiles of $B_T ( d )$ and $n_e ( d )$ as in Eq. (\ref{Bjet}), taking on average $\varphi = 45^{\circ}$ and $\Gamma = 10$. FSRQ jets are generally much longer than those of BL Lacs (in particular those of TeV emitting BL Lacs) -- reaching in the most extreme cases 1 Mpc -- and inflate giant ``radio lobes" in the external gas filled by a tenuous plasma. Radio polarimetric observations yield a good amount of information about the structure and the intensity of ${\bf B}$ in the radio lobes. Specifically, one gets a turbulent ${\bf B}$ which can be modeled as a domain-like structure with homogenous strength $B = 10 \, {\mu}{\rm G}$, coherence length $10 \, {\rm kpc}$ and random orientation of ${\bf B}$ in each domain. This magnetic field -- which is manifestly absent in BL Lacs -- provides another important region for $\gamma \to a$ conversions.

\section{Results}

The calculations have been performed following the standard procedure described in great detail e.g. in~\cite{dgr2011,1222}. Because of lack of space, here we merely report our results. We adopt $M = 10^{11} \, {\rm GeV}$ and $m < 10^{- 9} \, {\rm eV}$, not only for definiteness but also because these are the typical values for which some hints of VHE astrophysical effects of ALPs show up~\footnote{See the talk of G. Galanti at this Conference.} Basically, we display the the behaviour of $P_{\gamma \to a} (E)$ as a function of $E$ for a sample of benchmark values of $B_T (d_{{\rm ER}})$ and $d_{{\rm ER}}$ in the ranges considered above, discussing for clarity BL Lacs and FSRQs separately. Incidentally, owing to Eq. (\ref{Bjet}) the second of Eqs. (\ref{ETS}) becomes 
\begin{equation}
\label{ETS1}
E_H (d) = 2.10 \left(\frac{G}{B_T (d_{{\rm ER}})} \right) \left(\frac{d}{d_{{\rm ER}}} \right) \, {\rm GeV}~,
\end{equation}
from which we see that -- especially for BL Lacs -- at VHE energies the $\gamma \to a$ conversions occurs outside of the strong mixing regime for a sizable fraction of the jet.
\begin{figure}[h] 
\vskip -0.5 truecm
\center
\includegraphics[width=.6\textwidth]{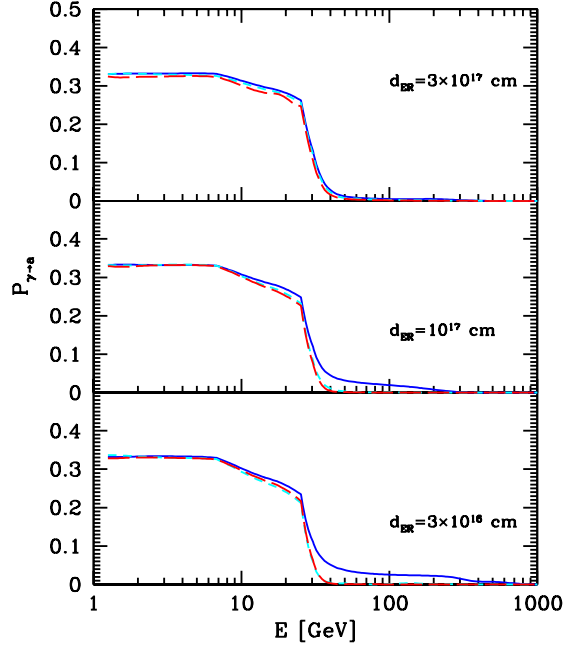} 
\vskip -2.3 truecm
\caption{Plot of $P_{\gamma \to a} (E)$ for a FSRQ including the conversion in the host galaxy and in the radio lobe. The different curves correspond to $B_T = 0.71 \, {\rm G}$ (solid blue), $2.13 \, {\rm G}$ (dashed cyan), $3.55 \, {\rm G}$ (long dashed, red). The three panels correspond to three values of the distance of the emitting region, namely $d_{{\rm ER}} =3 \cdot  10^{16} \, {\rm cm}$ (bottom), $10^{17} \, {\rm cm}$ (middle), $3 \cdot 10^{17} \, {\rm cm}$ (upper).} 
\label{fig1}
\end{figure}
Let us begin to address FSRQs. Their results are remarkably simple as shown in Fig.~\ref{fig1}. We choose as representative values $B_T (d_{{\rm ER}}) = 0.71, 2.13, 3.55 \, {\rm G}$ and we consider three different values $d_{{\rm ER}} =3 \cdot  10^{16} \, {\rm cm}$, $10^{17} \, {\rm cm}$, $3 \cdot 10^{17} \, {\rm cm}$ (see captions of Fig.~\ref{fig1}). At low energy $P_{\gamma\to a} (E)$ reaches in {\it all} cases the value $1/3$. Why? The explanation is remarkably simple. So long as the energy is small enough, the BLR absorption is pretty unimportant, they implying an efficient conversion. Further, the chaotic behaviour of ${\bf B}_T$ in the radio lobes leads to a nearly perfect ``equipartition" between the 2 degree of freedom of photons (two polarizations) and 1 degree of freedom of ALPs: this circumstance also erases any feature in $P_{\gamma \rightarrow a} (E)$ possibly induced by knots in the propagation along the jet. As the energy increases $P_{\gamma\to a} (E)$ monotonically decreases due to the enhanced optical depth in the BLR.

\begin{figure}[h] 
\vskip -0.5 truecm
\center
\includegraphics[width=.6\textwidth]{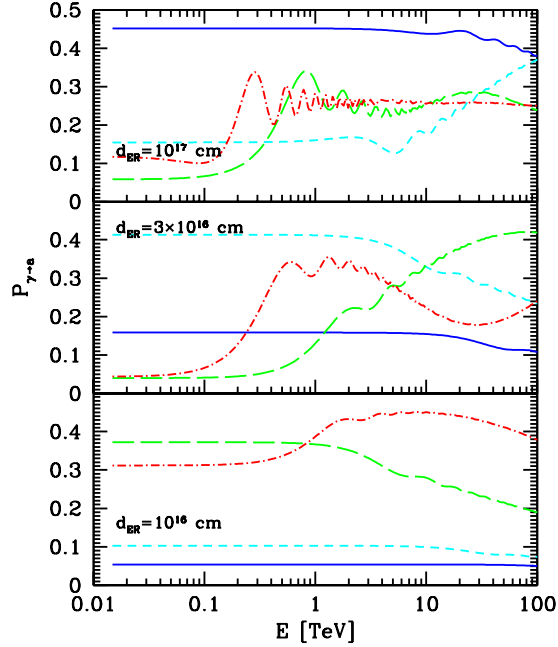} 
\vskip -2.3 truecm
\caption{Plot of $P_{\gamma \to a} (E)$ for a BL Lac including the host galaxy contribute. The different curves correspond to $B=0.1 \, {\rm G}$ (solid blue), $0.2 \, {\rm G}$ (dashed cyan), $0.5 \, {\rm G}$ (long dashed, green) and $1 \, {\rm G}$ (dot-dashed, red). The three panels correspond to three values of the distance of the emitting region, namely $d_{\rm ER} =10^{16} \, {\rm cm}$ (bottom), $3 \cdot 10^{16} \, {\rm cm}$ (middle), $10^{17} \, {\rm cm}$ (upper).} 
\label{fig2}
\end{figure}

Turning now our attention to the case of BL Lacs, we see from Fig.~\ref{fig2} that the situation becomes much more complex, in spite of their simpler structural properties. As it will become clear, the two facts are closely related. We take as benchmark values $B_T (d_{{\rm ER}}) = 0.1, 0.2, 0.5, 1 \, {\rm G}$ and we consider again there different values $d_{{\rm ER}} = 10^{16} \, {\rm cm}$, 
$3 \cdot 10^{16} \, {\rm cm}$, $10^{17} \, {\rm cm}$ (see captions of Fig.~\ref{fig2}). 

Let us start by considering the case $d_{{\rm ER}} = 10^{16}$ cm (bottom panel of Fig.~\ref{fig2}). In agreement with the previous discussion $P_{\gamma\rightarrow a} (E)$ smoothly decreases for energies above few hundreds of GeV, since the QED term becomes more and more important. Moreover, because the $a \gamma \gamma$ coupling constant goes like $B_T$, $P_{\gamma\rightarrow a} (E)$ increases with $B_T$. For $B=1$ G the probability around 1TeV is very close to the maximal conversion probability, $P_{\gamma\rightarrow a} (E) =0.5$ \cite{gala&ronca} (note that we assumed that the beam is unpolarized). At energies above 1TeV, the probability starts to decrease due to the greater importance of the QED term. 

Such a simple and intuitive picture breaks down for $d_{{\rm ER}} > 10^{16}$. Indeed, in such a situation (mid and top panels of  Fig.~\ref{fig2}) the curves show a complex behaviour, with the presence of multiple oscillations with different amplitude. Moreover, while the conversion probability monotonically increases with $B_T$ for $d_{{\rm ER}} = 10^{16}$ cm, this trend is not always preserved in the other two cases. Actually, for $d_{{\rm ER}} = 3 \cdot 10^{16}$ cm only the two cases with the lowest $B_T$ (0.1 G and 0.2 G) -- and for $d_{{\rm ER}} = 10^{17}$ cm only the case $B_T=0.1$ G -- exhibit a smooth behaviour. In all the other cases the conversion probability seems to follow an unpredictable pattern.

\begin{figure}[h] 
\vskip -0.5 truecm
\center
\includegraphics[width=.6\textwidth]{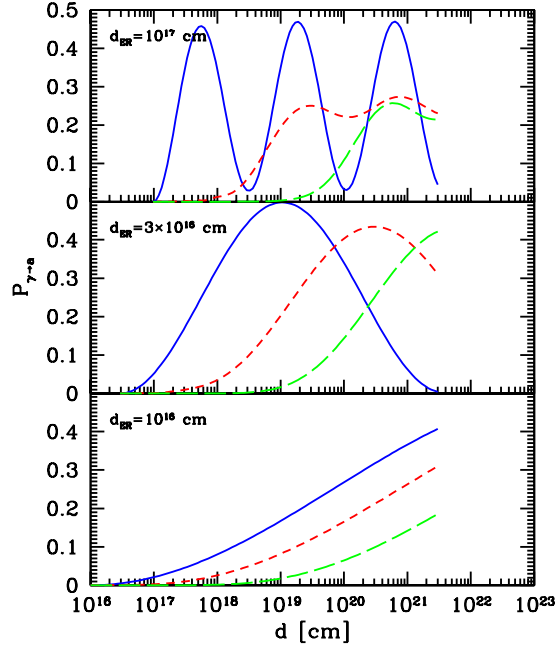} 
\vskip -2.3 truecm
\caption{Photon-axion conversion probability for a BL Lac as a function of the distance $d$ along the jet axis, for the case in which the emission region is located at $d_{\rm ER}=10^{16}$ cm (bottom), $3\times 10^{16}$ cm (middle) and $10^{17}$ cm (top). The different curves corresponds to $E=10$ GeV (solid blue), 500 GeV (dashed red), 10 TeV (long dashed, green). In all cases the magnetic field is $B=0.5$ G and the inverse coupling constant $M=10^{11}$ GeV.} 
\label{fig3}
\end{figure}

To gain a deeper insight into this issue, we report in Fig.~\ref{fig3} the behaviour of the conversion probability 
$P_{\gamma \rightarrow a} (E)$ as a function of the distance $d$ along the jet axis, for the same value $B_T = 0.5 \, {\rm G}$ of the magnetic field inside ${\cal R}_{\rm ER}$ for three different values of the energy: $E = 10 \, {\rm GeV}$, $E = 500 \, {\rm GeV}$ and $E = 10 \, {\rm TeV}$. Now, for $d_{\rm ER} = 10^{16} \, {\rm cm}$ (bottom panel) the probability increases monotonically with $d$. However, in the other two cases the plots show that $P_{\gamma \rightarrow a} (E)$ reaches a maximum and then decreases, describing one or more oscillations whose wavelengt depends on the energy $E$. Therefore oscillations at different energies are out of phase, ultimately determining the complex energy-dependence of the conversion probability. Clearly, in the case  $d_{\rm ER} = 10^{16} \, {\rm cm}$ the regular trend is due to the fact that we are merely seeing the onset of the first oscillation.

Our conclusions can be summarized as follows. For FSRQ the $\gamma \to a$ conversions are quite efficient for low enough energies where the internal absorption is irrelevant. Moreover, the presence of a chaotic ${\bf B}$ in the radio lobes brings about an equilibration among the degrees of freedom, thereby explaining why $P_{\gamma \to a} (E)$ is the same in the considered energy range for different values of $d_{{\rm ER}}$. For BL Lacs, instead, while generally $\gamma \to a$ conversions can be efficient -- especially for relatively large values of ${\bf B}$ in ${\cal R}_{\rm ER}$ -- it is impossible to assume that the conversion probability is maximal and energy-independent, as sometimes stated in the literature.


\begin{thebibliography}{99}   
\bibitem{alps} E. ~Masso, \emph{Lect. Notes Phys.} {\bf 741} (2008) 83; J. ~Jaeckel and A. ~Ringwald, \emph{Ann. Rev. Nucl. Part. Sci.} {\bf 60} (2010) 405. 
\bibitem{ringwald}  A. ~Ringwald, \emph{Phys. Dark Univ.} {\bf 1} (2012) 116; A. ~Ringwald, \emph{J. Phys. Conf. Ser.} {\bf 485} (2014) 012013.
\bibitem{cast} E. ~Arik {\it et al.} [CAST collaboration], \emph{JCAP} {\bf 02} (2009) 008.
\bibitem{HEW} W. ~Heisenberg and H. ~Euler, \emph{Z. Phys.} {\bf 98} (1936) 714; V. S. ~Weisskopf, \emph{K. Dan. Vidensk. Selsk. Mat. Fys. Medd.} {\bf 14} (1936) 6; J. ~Schwinger, \emph{Phys. Rev.} {\bf 82} (1951) 664.
\bibitem{rs1988} G. ~G. ~Raffelt and L. ~Stodolsky, \emph{Phys. Rev. D} {\bf 37} (1988) 1237.
\bibitem{urry} C.~M. Urry and P. Padovani, \emph{PASP}, {\bf 107} (1995), 803
\bibitem{darma} A. ~De ~Angelis, M. ~Roncadelli and O. ~Mansutti, \emph{Phys. Rev. D} {\bf 76} (2007) 121301; 
\bibitem{dgr2011} A. ~De ~Angelis, G. ~Galanti and M. ~Roncadelli, \emph{Phys. Rev. D} {\bf 84} (2011) 105030; (E) \emph{Phys. Rev. D} {\bf 87} (2013) 109903.
\bibitem{hooper2008} M. ~Simet, D. ~Hooper and P. ~D. ~Serpico, \emph{Phys. Rev. D} {\bf 77} (2008) 063001.
\bibitem{wb} D. ~Wouters and P. ~Brun, \emph{JCAP} {\bf 01} (2014) 014.      
\bibitem{harris} J. Harris and P. ~M. ~Chadwick, arXiv:1405.3227.
\bibitem{cluster} D. ~Horns {\it et al.}, \emph{Phys. Rev. D} {\bf 86} (2012) 075024.
\bibitem{prada} M. ~Sanchez-Cond\`e {\it et al.}, \emph{Phys. Rev. D} {\bf 79} (2009) 123511.
\bibitem{letter} F. ~Tavecchio, M. ~Roncadelli and G. ~Galanti, arXiv:1406.2303.
\bibitem{dermer14} C.~D. Dermer., \emph{proceedings for ``High Energy Astrophysics in Southern Africa,"} (2014), arXiv:1408.6453 
\bibitem{simul} A. Tchekhovskoy , R. Narayan, J.~C. McKinney, \emph{Mon. Not. R. Astron. Soc.}, 418 (2011), L79
\bibitem{lept} S. D. ~Bloom and A. P. ~Marscher, \emph{Astrophys. J.} {\bf 461} (1996) 657; F. ~Tavecchio, L. ~Maraschi and G. ~Ghisellini, \emph{Astrophys. J.} {\bf 509} (1998) 608.
\bibitem{hadr} F. A. ~Aharonian, \emph{Very High Energy Cosmic Gamma Radiation}, World Scientific, Singapore 2004.
\bibitem{gg10} G. Ghisellini {\it et al.}, \emph{Mon. Not. R. Astron. Soc.}, 402 (2010), 497
\bibitem{tavecchio2010} F. Tavecchio {\it et al.}, \emph{Mon. Not. R. Astron. Soc.} {\bf 401}, 1570 (2010).
\bibitem{bbr1984} M. C. Begelman, R. D. Blandford and M. J. Rees, \emph{Rev. Mod. Phys.} {\bf 56}, 255 (1984).
\bibitem{pudritz2011} R. E. Pudritz, M. J. Hardcastle and D. C. Gabuzda, \emph{Space Sci. Rev.} {\bf 169}, 27 (2012).
\bibitem{asada2008} K. Asada K. {\it et al.}, \emph{Astrophys. J.}, 682 (2008), 798 
\bibitem{gabuzda1999} D.~C. Gabuzda D.~C. {\it et al.}, \emph{Mon. Not. R. Astron. Soc.}, 307 (1999), 725
\bibitem{osullivan2009} S.~P. O'Sullivan and D.~C. Gabuzda, \emph{Mon. Not. R. Astron. Soc.}, 400 (2009), 26
\bibitem{giroletti2004} M. Giroletti {\it et al.}, \emph{Astrophys. J.}, 613 (2004), 752
\bibitem{1222} F. ~Tavecchio, M. ~Roncadelli, G. ~Galanti and G. ~Bonnoli, \emph{Phys. Rev. D} {\bf 86} (2012) 085036.
\bibitem{gala&ronca} G. Galanti and M. Roncadelli, arXiv:1305.2114.


\end{thebibliography}
\end{document}